\documentclass[sigconf]{acmart}
\usepackage{tabularx, multirow}

\AtBeginDocument{%
  }

\copyrightyear{2026}
\acmYear{2026}
\setcopyright{cc}
\setcctype{by}
\acmConference[MSR '26]{23rd International Conference on Mining Software Repositories}{April 13--14, 2026}{Rio de Janeiro, Brazil}
\acmBooktitle{23rd International Conference on Mining Software Repositories (MSR '26), April 13--14, 2026, Rio de Janeiro, Brazil}
\acmPrice{}
\acmDOI{10.1145/3793302.3793566}
\acmISBN{979-8-4007-2474-9/2026/04}

\begin{document}

\title{Understanding Dominant Themes in Reviewing Agentic AI-authored Code}

\author{Md. Asif Haider}
\email{asifh1@uci.edu}
\affiliation{%
  \institution{University of California, Irvine}
  \city{Irvine}
  \state{California}
  \country{USA}
}

\author{Thomas Zimmermann}
\email{tzimmer@uci.edu}
\affiliation{%
  \institution{University of California, Irvine}
  \city{Irvine}
  \state{California}
  \country{USA}
}



\renewcommand{\shortauthors}{Haider et al.}

\begin{abstract}
While prior work has examined the generation capabilities of Agentic AI systems, little is known about how reviewers respond to AI-authored code in practice. In this paper, we present a large-scale empirical study of code review dynamics in agent-generated PRs. Using a curated subset of the AIDev dataset, we analyze 19,450 inline review comments spanning 3,177 agent-authored PRs from real-world GitHub repositories. We first derive a taxonomy of 12 review comment themes using topic modeling combined with large language model (LLM)-assisted semantic clustering and consolidation. According to this taxonomy, we then investigate whether zero-shot prompts to LLM can reliably annotate review comments. Our evaluation against human annotations shows that open-source LLM achieves reasonably high exact match (78.63\%), macro F1 score (0.78), and substantial agreement with human annotators at the review comment level. At the PR level, the LLM also correctly identifies the dominant review theme with 78\% Top-1 accuracy and achieves an average Jaccard similarity of 0.76, indicating strong alignment with human judgments. Applying this annotation pipeline at scale, we find that apart from functional correctness and logical changes, reviews of agent-authored PRs predominantly focus on documentation gaps, refactoring needs, styling and formatting issues, with testing and security-related concerns. These findings suggest that while AI agents can accelerate code production, there remain gaps requiring targeted human review oversight. 

\end{abstract}





\maketitle

\section{Introduction}

Autonomous coding agents, also referred to as Agentic-AI Software Engineers, are now assisting humans in end-to-end development tasks, including coding, debugging, testing, submitting pull requests (PRs), acting on reviews, suggesting repairs, and generating documentation. This new paradigm of human-AI interaction faces some core challenges, such as establishing trust \cite{roychoudhury2025agentic}, as seen in human-driven workflows. Developers still hesitate to accept and merge AI-generated PRs, and the huge surge in the volume of code generated by AI agents just makes the situation worse. As reported in Hassan et al.~\cite{hassan2025agentic}, nearly 70\% of agent-authored PRs face longer review times, remain completely unreviewed, or get rejected soon, making the review process a major performance bottleneck. Hence, understanding the dynamics of AI-authored code reviews can fill this exposed gap on our path to enhancing human-AI collaboration. While datasets exist that capture how human-generated reviews are treated, finding large-scale empirical data on agentic-PRs has remained a challenge. Recent research in the mining software repository community presents datasets \cite{li2025rise, watanabe2025use} comprising PRs with rich metadata, authored by leading AI agents on the market. Using such data, we aim to address the following timely research questions:

\begin{itemize}
    \item \textbf{RQ1:} How accurately can LLMs annotate themes of code reviews commented on Agentic PRs?
    \item \textbf{RQ2:} What aspects/topics/themes are the most prevalent ones that receive attention while reviewing Agentic PRs?
    \item \textbf{RQ3:} What comment themes are dominant in successful vs rejected PRs?
\end{itemize}

Insights from these findings can be further leveraged to reduce failure rates, improve human-AI collaboration, and boost AI model training performance that prioritizes learning from mistakes in real-world projects. We provide our replication package via Zenodo~\cite{replicationpackage}.

\section{Dataset}


We utilize the AIDev \cite{li2025rise} dataset in our study. This is a large-scale dataset comprising Agentic-PRs from real-world GitHub projects. AIDev contains 932,791 Agentic-PRs authored by five most popular AI coding agents: OpenAI Codex, Devin, GitHub Copilot, Cursor, and Claude Code across 116,211 repositories involving 72,189 developers. Each PR in the dataset is linked to its corresponding repository and developer, along with additional metadata. The authors also released a curated subset of 33,596 Agentic-PRs from 2,807 repositories with more than 100 GitHub stars. This enriched subset provides review comments, commit-level diffs, event timelines, and related issues. For our analysis, we focus solely on this subset of AIDev. Specifically, we work on Zenodo version v2 of the data \cite{version2}. 

\begin{table}[t]
\centering
\tiny
\caption{Taxonomy of Review Comment Themes}
\label{tab:final_taxonomy}
\begin{tabular}{@{} l @{\hspace{1em}} >{\raggedright\arraybackslash}p{0.29\linewidth} >{\raggedright\arraybackslash}p{0.56\linewidth} @{}}
\toprule
\textbf{Tag} & \textbf{Thematic Category} & \textbf{Significant Keywords} \\
\midrule
\texttt{\textbf{secu}} & Security, Safety \& Reliability & vulnerability, security, tokens, safety, permissions, reliability, attack, defense \\
\midrule
\texttt{\textbf{test}} & Testing, Assertions \& Validation & test, assertion, expected, behavior, sql-checks, database, validation \\
\midrule
\texttt{\textbf{style}} & Styling \& Formatting Adjustments & line-level, text, formatting, rename, variable, attributes, readability, indentation, lint, consistency, style \\
\midrule
\texttt{docs} & Documentation, Logs \& Developer Guidance & documentation, api, guide, reference, section, markdown, changelog, typo, notes, logs \\
\midrule
\texttt{chore} & Dependency, Module \& Import Management & dependencies, module, import, export, package, sdk, optional-deps, resolution, cleanup \\
\midrule
\texttt{build} & Build, Configuration \& Default Behavior Management & build, permission, config, default, maven, gradle, cargo, script \\
\midrule
\texttt{ci} & CI/CD Workflow Management & workflow, ci, cd, versions, install, migration, github-actions, integration, deploy, travis \\
\midrule
\texttt{feat} & Core Implementation \& Feature Development & feature, new, implementation, logic, pattern, approach, functionality, helper, utility \\
\midrule
\texttt{refactor} & Code Restructure, Simplification \& Redundancy Elimination & unused, redundant, unnecessary, false-positive, refactor, dead-code, maintainability, simplify, restructure \\
\midrule
\texttt{\textbf{cmd}} & Command-Line Tools \& Developer Utilities & cli, commands, batch, python-tools, dev-utilities \\
\midrule
\texttt{\textbf{undo}} & Reverts, Rollbacks \& Change Rejection & revert, rollback, undo, restore, preserved, unchanged, rejection \\
\midrule
\texttt{\textbf{perf}} & Performance Improvement & memory, consumption, execution, speed, performance, monitor \\
\bottomrule
\end{tabular}
\end{table}

\section{RQ1: LLM as Review Theme Annotator}\label{method:rq1}

Here, we describe our approach to utilizing open-source LLMs for review comment topic categorization and evaluate LLM performance relative to human annotations. Since each PR may contain multiple review comments from reviewers, we first narrow our focus down to the \textit{pr\_review\_comments} table of the AIDev dataset \cite{version2}. This particular table contains inline code review comments with rich file-level context, including file path, diff hunks/code patches, and timestamps. To retrieve the PR level information and link it to the detailed review comments, we first run a join query combining the following tables: \textit{pull\_request}, \textit{pr\_reviews}, and \textit{pr\_review\_comments}. This leaves us with 19,450 review comments from 3,177 unique PRs.

\subsection{\textbf{Deriving a Taxonomy of Categories}} 

To derive a taxonomy for review comment categorization, we first take inspiration from the \textit{conventional commit specifications (CCS)} proposed by Zeng et al \cite{zeng2025first}, an approach adopted in recent literature for PR body content categorization \cite{watanabe2025use, li2025rise} as well. However, review comments differ from commit messages and PR bodies; hence, to identify review-specific textual signals in the review comment corpora, we further adopt a topic modeling-based approach to cluster potential themes with their corresponding keywords. In particular, we concatenate all review comment \textit{body} field entries for each PR into a document, and use transformer-based BERTopic modeling \cite{grootendorst2022bertopic} on the documents to identify unique potential themes and the most significant keywords associated with each hierarchical cluster.\\

Based on the keywords and 49 initial clusters, we then utilize ChatGPT (GPT-5)~\cite{gpt5} to gain a deeper understanding of the probable underlying definition of each cluster. To combine relevant clusters together, we also prompt ChatGPT to form groups of related clusters, keeping the unique keywords intact as much as possible. Thus, we identify 42 fine-grained topics and group them into 12 distinct thematic categories. Table \ref{tab:final_taxonomy} outlines the proposed set of \textbf{Thematic Category} with the most \textbf{Significant Keywords} associated with each topic. We also map each proposed theme to the aligning CCS category and term it as \textbf{Tag}. We chose not to allocate a different category for \textit{\textbf{fix}}, since fixing an issue can fall under security, styling, or even testing-related changes. Rather, we introduce a new category titled \textit{\textbf{undo}}, capturing highly abundant reviews like file restoration and simple undo/revert operations for a given code patch. We also separate command-line tools and developer-utility-specific suggestions from the \textit{\textbf{build}}/\textit{\textbf{ci}} categories, and call it \textit{\textbf{cmd}}. Finally, although our topic analysis could not identify keywords unique enough to map to the \textit{\textbf{perf}} tag, we chose to preserve this category based on frequent keywords found in the literature \cite{watanabe2025use}. 


\begin{figure*}[t]
  \centering
\includegraphics[width=\linewidth]{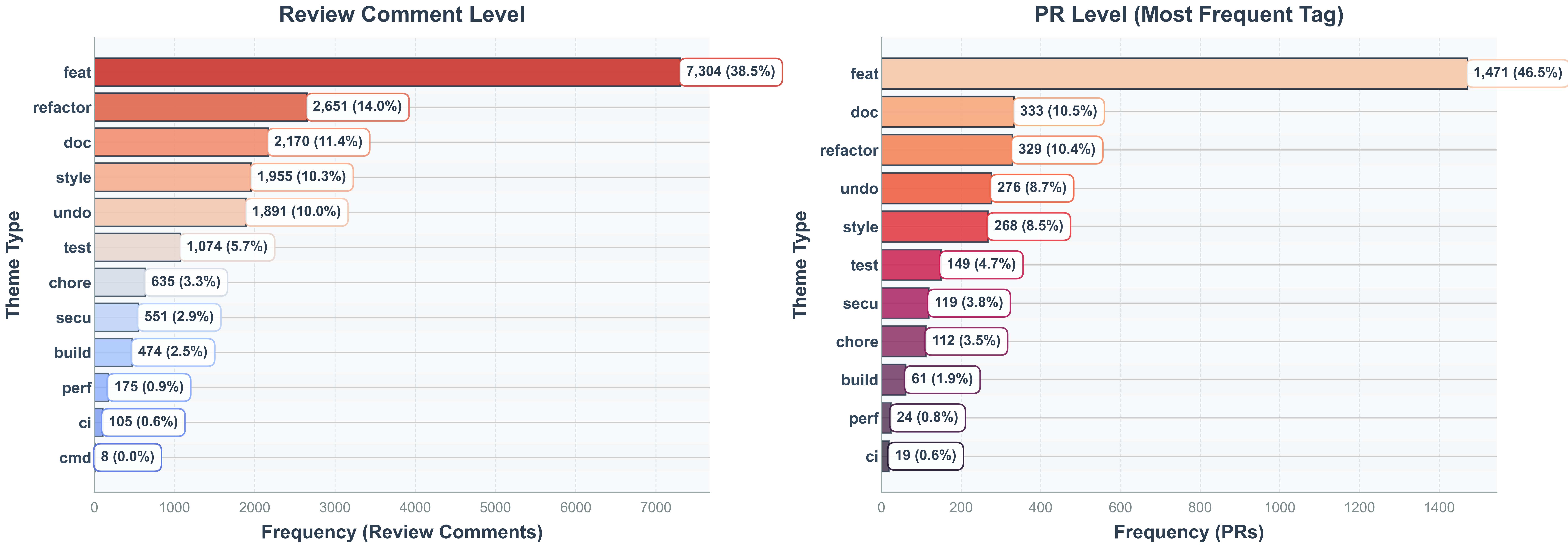}
  \caption{Distribution of review themes across 19,007 individual review comments and 3,162 PRs}
\label{fig:barplot}
\end{figure*}

\subsection{\textbf{Annotation Performance: LLM vs Human}}\label{prompting-info}
Based on our thematic analysis of almost 19.5k review comments, we derive 12 categories, their descriptions, and the most significant keywords. Using this taxonomy information as the input, we then prompt an open-source LLM to generate the most suitable tag for each review comment. We chose \textbf{Gemma 3:12B} model via ollama \cite{gemma312} after a primary experiment with a pool of LLMs of varying size, including Llama 3.2:3B, Llama 3.1:8B, and Deepseek-R1:14B variants. We omitted proprietary LLMs to avoid incurring unnecessary costs and to support reproducibility. The prompt template with topic clusters is available in our replication package \cite{replicationpackage}. For our preliminary annotation experiment, we randomly sampled 100 PRs from the initial subset, which covered 571 unique review comments. We manually annotate the same test set of PRs and compare LLM performance across two layers. First, we consider the \emph{review comment level} tags and report the following single-tag metrics:

\begin{itemize}
    \item \textbf{Exact Match (EM):} Also known as \textit{Accuracy}, in this case, EM refers to the percentage of correctly tagged review comments by LLMs, when compared to the human-annotated ones. This indicates the LLM's overall accuracy.
    \item \textbf{Precision, Recall and F1 Score (Macro):} The harmonic mean of macro precision and recall, used to determine if the LLM performs well across all types of comments or not. 
    \item \textbf{Cohen's Kappa ($\kappa$):} Considering the LLM as an independent annotator, we measure its agreement level with a human annotator to account for mere chance. A high \textbf{$\kappa$} is an indicator of LLMs being a reliable proxy for human annotators.  
\end{itemize}

We also consider the \emph{PR level} evaluation to assess how well LLM can identify the dominant themes in a PR containing review comments of multiple categories. To achieve so, we keep up to 3 of the most frequent tags in each PR, and report the following metrics:

\begin{itemize}
    \item \textbf{Jaccard Similarity:} This set overlap metric essentially ignores the order of the tags and simply asks what the overlap between the Top-3 tags annotated by humans and LLM is. It is useful to identify if the LLM can capture the general \textit{cloud} of themes. Mathematically, $J(A, B) = \frac{|A \cap B|}{|A \cup B|}$ 
    \item \textbf{Top-1 Accuracy:} We also check if the LLM could correctly identify the single most frequent tag in the PR, which essentially attempts to confirm whether the LLM understands the primary theme that drives the review discussion.
\end{itemize}




At the \emph{review comment level}, the model achieves an Exact Match of 0.7863, indicating that a substantial majority of individual review comments were assigned the correct label. The macro-averaged precision (0.8691) and recall (0.7308) suggest that the model maintains a good balance between avoiding false positives and accurately identifying instances across all classes. Similarly, the resulting macro F1 score of 0.7756 reflects consistent performance across all classes, even under class imbalance. Moreover, Cohen’s $\kappa$ value of 0.7348 indicates substantial agreement beyond chance, underlining the reliability of the model’s annotations. At the \emph{PR level}, the accuracy remains similar, with a Top-1 accuracy of 0.78, demonstrating that the model can effectively infer the dominant review label for an entire pull request. Macro precision (0.9236), recall (0.8730), and F1 score (0.8819) all exhibit considerable improvement over their granular counterparts. The average Jaccard Similarity of 0.8142 indicates a high degree of overlap between the predicted and ground-truth label sets, suggesting that the model captures the semantic intent of review comments even when exact matches are not fully achieved. 

\section{RQ2: Top Themes in Agentic PR Review}

Continuing with the same prompting technique mentioned in section \ref{prompting-info}, we identified the review theme categories at both the review comment and PR levels. We started with our full experiment set of 19,450 review comments from 3,177 distinct PRs, as outlined in section \ref{method:rq1}. However, we excluded invalid LLM responses from our further analysis that either do not comply with the desired, prompted output structure or do not generate outputs from the defined pool of 12 tags, or are simply too large to be processed. This ultimately leaves us with 19,007 review comments across 3,162 PRs. \\


\begin{figure*}[t]
  \centering
\includegraphics[width=\linewidth]{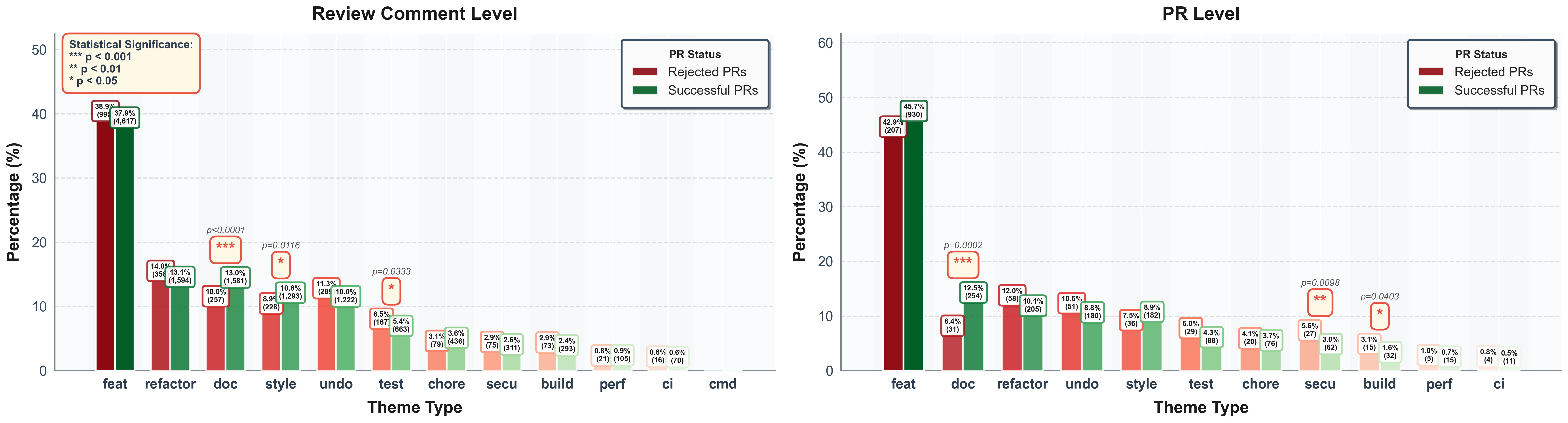}
  \caption{Percentage of review themes across 483 rejected and 2035 accepted PRs}
\label{fig:rejected_barplot}
\end{figure*}

Figure \ref{fig:barplot} demonstrates the distribution of review themes at both the review comment and PR level. Suggestions related to functional correctness, logical changes, and feature development \textit{(\textbf{feat})} remain the single most frequent review themes at both granular (38.5\%) and coarse-level (46.5\%). Restructuring, simplification, and redundancy elimination (\textbf{\textit{refactor}}, 14\% and 10.4\%), and documentation guides with changelogs (\textbf{\textit{doc}}, 11.4\% and 10.5\%) occupy the next two places respectively. Styling and formatting adjustments to ensure better readability and consistency (\textbf{\textit{style}}, 10.3\% and 8.5\%) and straightforward reverts, rollbacks, and abort messages (\textbf{\textit{undo}}, 10\% and 8.7\%) comprise the next majority of dominant review comments. Testing concerns with assertion and validation checks (\textbf{\textit{test}}) consistently remain the fifth most important aspect of agentic PR review, accounting for more than 5\% of the comments on average. Dependency, module, and import management, version controlling with codebase-wide cleanup comments appear next in the list (\textbf{\textit{chore}}, 3.3\% and 3.5\%). Security concerns (\textbf{\textit{secu}}, 2.9\% and 3.8\%), such as token management, SQL injection attacks, and comments related to vulnerability, safety, and reliability assessment, are also important concerns in agentic code review. Finally, the least frequent ones appear to be related to build, configuration, scripting, and default behavior management (\textbf{\textit{build}}, 2.5\% and 1.9\%), and performance (memory, runtime, latency) improvement recommendations (\textbf{\textit{perf}}) at nearly 1\%. CI/CD workflow suggestions (\textbf{\textit{ci}}) are provided in only 0.6\% of cases. It is interesting to note that command-line tools and utilities-related issues (\textbf{\textit{cmd}}) appear rarely in individual review comments; however, they never become the dominant tag in any of the analyzed PRs, suggesting this category is a weak one overall. In fact, it can be paired up with \textbf{\textit{build}} or \textbf{\textit{ci}} depending on the context.

\section{RQ3: Accepted vs Rejected PR Review Themes}

To answer this question, we first define unsuccessful/rejected (R) PRs as PRs with valid \textit{closed\_at} entries but null values in the \textit{merged\_at} column. Thus, we ended up with a subset of 2,558 review comments across 483 PRs. Similarly, we extracted successfully accepted (A) PRs with both valid \textit{closed\_at} and \textit{merged\_at} entries, yielding a total of 12,191 review comments across 2,035 PRs. Figure \ref{fig:rejected_barplot} shows the results from our further analysis on these two subsets. While features and functional correctness \textbf{\textit{(feat)}} remain the primary concern in both cases, some statistically significant shift of focus emerges when PRs are ultimately rejected. Chi-Square tests reveal that at the granular level, documentation \textbf{\textit{(doc)}} (10.05\% (R) vs 12.97\% (A); $p < 0.001$) and styling concerns \textbf{\textit{(style)}} (8.91\% (R) vs 10.61\% (A); $p < 0.05$) are significantly more common in successful PRs; suggesting that these are viewed as constructive hurdles, and the reviewers are willing to invest time in refinement given the core logic is sound already. However, testing \textbf{\textit{(test)}} issues are significantly more prevalent in rejected PRs (6.53\% (R) vs 5.44\% (A); $p < 0.05$), underscoring their potential as indicators of rejection. At the PR level, the potential drivers of failure become even more distinct. Security concerns \textbf{\textit{(secu)}} emerge as a dominant theme in rejected PRs (5.59\% (R) vs 3.05\% (A); $p < 0.01$). Configuration and build issues \textbf{\textit{(build)}} are also likely to trigger rejections (3.11\% (R) vs 1.57\% (A); $p < 0.05$). Again, \textbf{\textit{doc}} shows a strong prevalence in accepted PRs (6.42\% (R) vs 12.48\% (A); $p < 0.001$). Even though not statistically significant enough, \textbf{\textit{undo}} comments are notably more prevalent in rejected PRs at both levels, highlighting the high amount of unnecessary reverts committed by agents that reviewers eventually reject. 


\section{Limitations}
Despite substantial agreement with humans, LLM-based labeling may still misinterpret or confuse highly nuanced technical terms and project-specific keywords, potentially impacting the distribution of themes. Our findings are based solely on the AIDev dataset \cite{li2025rise}. While it compiles rich data from popular repositories, the results may not generalize well to closed-source enterprise environments under varied review standards. One author primarily annotated the validation set; however, multiple annotators could establish better inter-rater reliability. We categorized PRs as rejected based on the absence of a valid merge timestamp. This can potentially include PRs that were closed due to repository maintenance and archival issues, rather than technical shortcomings. We also acknowledge the small size of the validation set, which may compromise the robustness of our results. We used the Chi-Square test for individual significance testing; however, some individual categories (like \textbf{\textit{undo}}) were borderline ($p\approx0.057$). An increased sample size could help clarify such marginal trends. Finally, correcting for multiple comparisons would reduce the risk of Type I statistical errors.


\section{Related Works}
Several studies over the past years have explored techniques to improve the quality and automation of code review. These efforts have gradually progressed from Information Retrieval-based methods \cite{hong2022commentfinder}, to pretrained Deep Learning models \cite{li2022automating, tufano2022using, li2022auger,ochodek}, and most recently to solutions leveraging LLMs \cite{pornprasit2024fine, lu2023llama,haider2024prompting,rasheed2024ai,vijayvergiya2024ai}. However, these solutions have drawbacks in the context of agentic software engineering. First, automation approaches often lack a solid grounding in real-world development workflows and fail to account for the unique dynamics of an end-to-end agentic pipeline. Practical performance is also not particularly promising, even when keeping agentic variants aside. Cihan et al. \cite{cihan2025automated} attempted to measure practitioners' perceived usefulness of code review automation in industry settings, and found that longer PR closure times and irrelevant comments were common. Previous research \cite{davila2025fine, ochodek} has also attempted to establish taxonomies of code review comments. While serving as good starting points, they are either based on a single repository or focus only on human-centric review workflows.

\section{Ethical Implications}
We analyze publicly available GitHub repositories under permissive licenses (MIT or Apache 2.0). We also do not identify any information about individual developers or projects. Our findings should be used to enhance the agent's ability to self-correct before burdening the human reviewer, rather than fully replacing human oversight.

\section{Conclusion and Future Work}

In this work, we study the review dynamics of Agentic AI-authored code by first deriving a comprehensive taxonomy of 12 thematic categories using topic modeling and semantic clustering. Our LLM-powered annotation could reliably identify review themes in PRs, and further analysis provides actionable insights for reviewers and practitioners. Our experimental results suggest that \emph{documentation and styling aspects can be constructive obstacles in code review; however, problems in testing, security, and build configuration are more likely to block a successful code merge}. Future agentic architectures must move beyond generative capabilities and incorporate stronger internal validation loops to overcome these merge hurdles. At the model level, reducing agentic \textit{noise}- unnecessary changes not contributing to the PR's goal- can reduce reviewer fatigue and improve the perceived reliability of AI teammates. Next, we plan to improve annotation performance, refine the rejection criteria, and consider non-technical failure reasons as well. Another potential future line of work can be leveraging the identified failure patterns (e.g., security and testing) to create targeted fine-tuning datasets that support better model training for automated review and repair.

\newpage

\bibliographystyle{ACM-Reference-Format}
\bibliography{sample-base}

\end{document}